\documentstyle[12pt]{article}
\newcommand{\bea}{\begin{eqnarray}}
\newcommand{\eea}{\end{eqnarray}}
\newcommand{\be}{\begin{equation}}
\newcommand{\ee}{\end{equation}}
\newcommand{\vs}[1]{\vspace{#1 mm}}

\renewcommand{\a}{\alpha}
\renewcommand{\b}{\beta}
\renewcommand{\c}{\gamma}
\renewcommand{\d}{\delta}

\newcommand{\dsl}{\pa \kern-0.5em /}

\newcommand{\pa}{\partial}

\newcommand{\nn}{\nonumber\\}

\begin{document}
\topmargin 0pt
\oddsidemargin 0mm



\begin{flushright}

USTC-ICTS-05-8\\


hep-th/0508045\\


\end{flushright}

\vspace{2mm}

\begin{center}

{\Large \bf

Fundamental strings and NS5-branes from unstable\\
D-branes in supergravity}

\vs{6}

{\large J. X. Lu$^a$\footnote{E-mail: jxlu@ustc.edu.cn}
 and Shibaji Roy$^b$\footnote{E-mail: shibaji.roy@saha.ac.in}}

 \vspace{4mm}

{\em

 $^a$ Interdisciplinary Center for Theoretical Study\\

 University of Science and Technology of China, Hefei, Anhui
 230026, China\\
 and \\
 Center for Mathematics and Theoretical Physics\\
 Institute for Advanced Study\\
 University of Science and Technology of China, Shanghai 201315, China\\

and\\

Interdisciplinary Center of Theoretical Studies\\

Chinese Academy of Sciences, Beijing 100080, China\\




\vs{4}

 $^b$ Saha Institute of Nuclear Physics,

 1/AF Bidhannagar, Calcutta-700 064, India}

\end{center}

\begin{abstract}
By using the non-supersymmetric $p$-brane solutions delocalized in arbitrary
number of transverse directions in type II supergravities, we show how they can be
regarded as interpolating solutions between unstable D$p$-branes (a non-BPS
D-brane or a pair of coincident D-brane-antiD-brane) and
fundamental strings and also between unstable D$p$-branes and NS5-branes. We also
show that some of these solutions can be regarded as interpolating solutions
between NS5/$\overline{\rm NS}$5 and D$p$-branes (for $p \leq 5$). This gives
a closed string description of the tachyon condensation and lends
support to the conjecture that the open string theory on unstable D-branes
at the tachyonic vacuum has soliton solutions describing not only the
lower dimensional BPS D-branes, but also the fundamental strings as well as
the NS5-branes.
\end{abstract}
\newpage










\renewcommand{\a}{\alpha}

\renewcommand{\b}{\beta}

\renewcommand{\c}{\gamma}

\renewcommand{\d}{\delta}













\topmargin 0pt

\oddsidemargin 0mm

A non-BPS D-brane or a pair of coincident brane-antibrane system in type II
string theories is
unstable which is manifested by the presence of open string tachyon on their
respective world-volumes
\cite{Sen:1998sm, Sen:1999mg}.
Because of this, these unstable D-branes decay and the decay occurs through
a process called tachyon condensation. The tachyon condensation
has been understood from the open string point of view both in the boundary CFT
approach \cite{Callan:1994ub,Polchinski:1994my,Recknagel:1998ih,Sen:1999mg} and
the string field theory approach
\cite{Sen:1999nx,Berkovits:2000hf}. In particular,
it has been
conjectured that the tachyonic vacuum in these theories describes
the closed string vacuum without any brane and the various solitonic solutions
representing the lower dimensional BPS D-branes, the fundamental strings as well as
the Neveu-Schwarz 5-branes \cite{Sen:2000kd}. This is what is expected
and evidences have been given for the lower
dimensional BPS D-branes in \cite{Sen:1999mh} (the appearance of these objects can also
be understood from K-theory \cite{Witten:1998cd}) and for the fundamental strings
in \cite{Sen:2000kd,Yi:1999hd}
from the open string point of view.

The appearance of the various decay products at the tachyonic
vacuum from the D$p$-$\overline{\rm D}p$ (we denote an anti-brane
with a bar) or from a non-BPS D$(p-1)$-brane can be understood as
follows \cite{Yi:1999hd,Bergman:2000xf}. For definiteness we will
discuss the case of D$p$-$\overline{\rm D}p$-brane and briefly
mention about non-BPS D$(p-1)$-branes.  The world-volume theory on
a pair of D$p$-$\overline{\rm D}p$ brane is a gauge theory with
gauge group $U(1) \times U(1)$ and has a complex tachyon of charge
(1, $-$1). (For non-BPS D$(p-1)$-brane the gauge group is $Z_2
\times U(1)$ with a chargeless, real tachyon. The removal of
$U(1)$ gauge degrees of freedom here can be achieved through
confinement which can be discussed similarly as for the case of
D$p$-$\overline{\rm D}p$-brane.) As the tachyon condenses, its
phase can acquire a non-trivial winding number on a finite energy
vortex solution of this theory which can be identified as BPS
D$(p-2)$-brane. In this process the relative $U(1)$ gauge group is
broken and the relative gauge field $(A_1-A_2)$ gets removed by
the Higgs mechanism. The overall gauge field $(A_1+A_2)$ under
which the tachyon is neutral gets also removed through
confinement, which can also be understood by the dual Higgs
mechanism. The only remnant of the gauge field is the confined
electric flux string which can be identified as the fundamental
string \cite{Yi:1999hd,Bergman:2000xf}. This phenomenon can be
understood by open D$(p-2)$-branes stretched between
D$p$-$\overline{\rm D}p$-brane. The open D$(p-2)$-brane will
induce a $(p-3)$ dimensional tachyonic object charged under the
relative two $(p-2)$-forms $(A_1^{[p-2]} - A_2^{[p-2]})$ on the
world-volume of D$p$-$\overline{\rm D}p$-brane. The dual of the
corresponding field-strength is related with the electric flux
associated with the overall gauge field $(A_1+A_2)$. So, the
$(p-3)$-dimensional tachyonic object is magnetically charged under
$(A_1+A_2)$ and after its condensation the overall $U(1)$ is
removed by the dual Higgs mechanism \cite{Yi:1999hd} (a different
mechanism, as opposed to $p>2$, has been suggested
\cite{Bergman:2000xf} for $p\leq 2$ cases) giving a confined
electric flux string identified as F-string.

We now come to NS5-brane. The NS5-$\overline{\rm NS}5$
decaying into D$p$ branes with $p \le 5$ can be deduced either from
the result  NS5-$\overline{\rm NS}5$ decaying into D2
\cite{Yi:1999hd} and by applying T-dualities subsequently along
the NS5 directions or from the D5-$\overline{\rm D}5$ decaying into D3
and by applying S-duality and subsequent T-dualities. On
the other hand, D$p$-$\overline{\rm D}p$ decaying into NS5 is less
obvious but can still be deduced partially from the above. We
mentioned NS5-$\overline{\rm NS}5$ decays into D5 and S-dual
of this gives D5-$\overline{\rm D}$5 decaying into NS5. Applying
T-dualities along the NS5 directions gives D$p$-$\overline{\rm D}p$
decaying into NS5 for $p \le 5$. Now in order to see the process
D6-$\overline{\rm D}$6 decaying into NS5, we can start with D0-$\overline{\rm D}$0
decaying into F-string \cite{Bergman:2000xf} and apply T- and S-duality
successively on it. To be precise, if the F-string above lies along
$x^1$-direction, then the application of (T$_1$ST$_{34}$ST$_{56}$ST$_{234}$)
from the right on both sides of D0-$\overline{\rm D}$0
decaying into F, where the subscript in T refers
to application of T-duality along those directions and S refers to
S-duality, will give D6-$\overline{\rm D}$6 decaying into NS5.
Some
relevant discussions can also be found in \cite{Eyras:1999at,Houart:1999bi}.

In this paper we will give a closed string or supergravity description of
the appearance of F-strings and NS5-branes from D$p$-$\overline{\rm D}p$-brane
or non-BPS D$p$-brane described above. It should be pointed out that
in the supergravity picture there is no explicit tachyon field, but as
it was argued in \cite{Brax:2000cf,Lu:2004dp} tachyon vev can appear as
parameters labelling the
supergravity solution. Indeed the non-susy $p$-brane solutions
\cite{Zhou:1999nm,Lu:2004ms}
which in some special cases can be argued to represent D$p$-$\overline{\rm D}p$-brane
system or non-BPS D$p$-brane contain some parameters. By using their supersymmetric
reduction we have argued in \cite{Lu:2004dp}, how these parameters are related to the
microscopic physical parameters as well as the tachyon vev to correctly describe
the tachyon condensation process. In a different approach we have shown earlier
\cite{Lu:2004xi}, how by delocalizing the non-susy D$p$-brane solution in one transverse
direction, we can regard it (a) as an interpolating solution between non-BPS D$(p+1)$
brane and a BPS D$p$-brane similar to the picture of tachyon condensation on the
kink solution \cite{Sen:1999mg} and (b) as an interpolating solution between
non-BPS D$(p+1)$-brane
and the supergravity description of tachyon matter \cite{Ohta:2002ac}
similar to the rolling tachyon
picture \cite{Sen:2004nf} in the open string description. For case (a)
the delocalized direction was
space-like and for (b) the delocalized direction was time-like. This result gave us
confidence to believe that the supergravity solution indeed correctly represent
the D$p$-$\overline{\rm D}p$-brane system or non-BPS D$p$-brane. We have
further shown in \cite{Lu:2005ju}, by delocalizing the non-susy D$p$-brane to two spatial
transverse directions, how this solution can be regarded as an interpolating solution
between D$(p+2)$-$\overline{\rm D}(p+2)$, non-BPS D$(p+1)$ and BPS D$p$-brane
giving the complete descent relation of Sen \cite{Sen:1999mg}
under tachyonic kink and vortex solutions.
Since we have already described how the lower dimensional D-branes can be
obtained from D$p$-$\overline{\rm D}p$ and non-BPS D$p$ in our earlier works,
in this paper we will show how the other closed string objects like F-strings and
NS5-branes can be obtained from D$p$-$\overline{\rm D}p$ (or non-BPS D$p$).
We will also show how D$p$-branes can be obtained from NS5-${\overline{\rm NS}}$5.
This gives a closed string description of tachyon condensation on unstable D-brane
and gives more evidence to the conjecture that tachyonic vacuum is a closed string
vacuum consisting not only of lower dimensional D-branes, but also of F-strings
and NS5-branes.

In order to show various processes described above from supergravity we first
construct the non-supersymmetric $p$-brane solutions delocalized in $q$ spatial
directions in arbitrary space-time dimensions $d$. The  supergravity
action we consider is,
\be S = \int d^dx \sqrt{-g}
\left[R - \frac{1}{2}
\partial_\mu \phi \partial^\mu \phi - \frac{1}{2\cdot (d-p-2)!}
e^{a\phi} F_{[d-p-2]}^2\right] \ee 
where $g_{\mu\nu}$, with $\mu,\,\nu
= 0,1,\ldots,d-1$, is the metric and $g={\rm det}(g_{\mu\nu})$,
$R$ is the scalar curvature, $\phi$ is the dilaton, $F_{[d-p-2]}$ is
the field strength of a $(d-p-3)$-form gauge field and $a$
is the dilaton coupling. The action (1) represents the bosonic
sector of the low energy effective action of string/M theory
dimensionally reduced to $d$-dimensions. Now in order to obtain
the delocalized solutions in $q$ transverse directions, we have to
solve the equations of motion from (1) with the following ansatz
for the metric and the $(d-p-2)$-form field strength, \bea ds^2 &=&
e^{2A(r)}\left(dr^2 + r^2 d\Omega_{d-p-q-2}^2\right) + e^{2B(r)}
\left( -dt^2 + \sum_{i=1}^p dx_i^2 \right) + \sum_{i=2}^{q+1}e^{2C_{i-1}(r)} 
dx_{p+i-1}^2\nn F_{[d-p-2]} &=& b\,\, {\rm
Vol}(\Omega_{d-p-q-2}) \wedge dx_{p+1} \ldots \wedge dx_{p+q} \eea 
In the
above $r = (x_{p+q+1}^2 + \cdots + x_{d-1}^2)^{1/2}$,
$d\Omega_{d-p-q-2}^2$ is the line element of a unit
$(d-p-q-2)$-dimensional sphere, Vol($\Omega_{d-p-q-2}$) is its
volume-form and $b$ is the magnetic charge parameter. The
solutions (2) represent magnetically charged $p$-brane solutions
delocalized in transverse $x_{p+1}, x_{p+2} \ldots, x_{p+q}$, directions. The
equations of motion will be solved with the following gauge
condition, 
\be 
(p+1) B(r) + (d-p-q-2) A(r) + \sum_{i=2}^{q+1}C_{i-1}(r) = \ln G(r)
\ee 
Note that as $G(r) \to 1$, the above condition reduces to the
extremality or the supersymmetry condition. As
mentioned in \cite{Lu:2004ms}, the consistency of the equations of
motion dictates that the non-extremality function $G(r)$ can take
three different forms and we will need only one of them for our
purpose which is, 
\be G(r) = 1-\frac{\omega^{2(d-p-q-3)}}{r^{2(d-p-q-3)}}
= \left(1+\frac{\omega^{d-p-q-3}}
{r^{d-p-q-3}}\right)\left(1-\frac{\omega^{d-p-q-3}} {r^{d-p-q-3}}\right) =
H(r) \tilde{H}(r) \ee 
By solving the equations of motion following from (1), we can obtain all the 
functions $A(r)$, $B(r)$ and $C_1(r), \ldots, C_q(r)$ appearing in the metric
subject to some constraints. The solutions therefore take the forms,
\bea 
ds^2 &=&
F^{\frac{4(p+1)}{(d-p-3)\chi}} (H{\tilde {H}})^{\frac{2}{d-p-q-3}}
\left(\frac{H}{\tilde
H}\right)^{\frac{-2\sum_{i=2}^{q+1}\delta_i}{d-p-q-3}}\left(dr^2 + r^2
d\Omega_{d-p-q-2}^2\right)\nn
& & \qquad\qquad + F^{-\frac{4}{\chi}}\left(-dt^2 +
\sum_{i=1}^p dx_i^2 \right) +
F^{\frac{4(p+1)}{(d-p-3)\chi}} \sum_{i=2}^{q+1}\left(\frac{H}{\tilde
H}\right)^{2\d_i} dx_{p+i-1}^2 \nn 
e^{2\phi} &=& F^{-\frac{4a(d-2)}{(d-p-3)\chi}} \left(\frac{H}{\tilde
{H}}\right)^{2\delta_1},\qquad F_{[d-p-2]}\,\,\, =\,\,\, b\,\,{\rm
Vol}(\Omega_{d-p-q-2})\wedge dx_{p+1} \ldots \wedge dx_{p+q} 
\eea 
These are
magnetically charged, non-supersymmetric $p$-brane
solutions delocalized in $q$ directions and the corresponding electrically 
charged solutions can
be obtained by simply replacing $F \to e^{-a\phi} \ast F$. The field-strength
for the electrical solution takes the form (which follows from (5))
\be
F_{[p+2]} = \frac{\sinh2\theta}{2} d\left(\frac{C}{F}\right)\wedge dt \wedge
dx_1 \ldots \wedge dx_p
\ee
The functions $F$ and $C$ appearing in the above solutions are defined as
\bea
F &=& \cosh^2\theta \left(\frac{H}{\tilde H}\right)^\a
- \sinh^2\theta \left(\frac{\tilde H}{H}\right)^\b\nn
C &=& \left(\frac{H}{\tilde H}\right)^\a - \left(\frac{\tilde H}{H}\right)^\b
\eea 
In the solutions (5) and (6) above there are $(q+5)$ integration constants and they are
$\omega$, $\theta$, $\alpha$, $\beta$, $\d_1, \ldots, \d_{q+1}$. $b$ is the charge
parameter. However, not all the parameters are independent. From consistency
of the equations of motion we obtain three relations among the parameters and they
are
\bea
&&\alpha-\beta \,\,\,= \,\,\,a\delta_1\\
&&\frac{1}{2} \delta_1^2 + \frac{2\alpha(\alpha-a\delta_1)(d-2)}{\chi(d-p-3)} +
\frac{2 \sum_{i<j=2}^{q+1}\delta_i\delta_j}{d-p-q-3} \,\,\,=\,\,\, 
\left(1-\sum_{i=2}^{q+1}\delta_i^2\right)
\frac{d-p-q-2}{d-p-q-3}\\
&&b\,\,\, =\,\,\, \sqrt{\frac{4(d-2)}{(d-p-3)\chi}} (d-p-q-3)
\omega^{d-p-q-3} (\alpha+\beta) \sinh2\theta 
\eea 
where $\chi$ in the above is defined as $\chi=2(p+1) + a^2(d-2)/(d-p-3)$ and the dilaton
coupling is given by $a^2 = 4 - 2(p+1)(d-p-3)/(d-2)$ for the supergravities with
maximal supersymmetry (the case we are considering). So, the number of independent 
parameters in the above solutions is $(q+3)$.
  
For our purpose of showing the appearance of F-strings from the D$p$-$\overline{\rm D}p$
system 
we here write the non-susy F-string solutions
delocalized
in $(p-1)$ transverse directions from the general solutions (5) and (6) by putting
$d=10$, $p=1$, $q=p-1$, $a=1$ (which implies $\chi=16/3$) as,
\bea
ds^2 &=&
F^{\frac{1}{4}} (H{\tilde {H}})^{\frac{2}{7-p}}
\left(\frac{H}{\tilde H}\right)^{-(2\sum_{i=2}^p \delta_i)/(7-p)}\left(dr^2 + r^2
d\Omega_{8-p}^2\right)\nn
&& + F^{-\frac{3}{4}}\left(-dt^2 + dx_1^2
\right) + F^{\frac{1}{4}}\sum_{i=2}^p
\left(\frac{H}{\tilde H}\right)^{2\d_i} dx_i^2\nn
e^{2\phi} &=& F^{-1}
\left(\frac{H}{\tilde {H}}\right)^{2\delta_1},
\quad B^{(2)}\,\,=\,\, \frac{\sinh2\theta}{2}\left(\frac{C}{F}\right)dt \wedge
dx_1
\eea
with the parameter relation
\be
\frac{1}{2}\d_1^2 + \frac{1}{2} \a(\a-\d_1) + \frac{2\sum_{i<j=2}^p \d_i \d_j}{7-p}
= \left(1-\sum_{i=2}^p \d_i^2\right) \frac{8-p}{7-p}
\ee
The functions $F$ and $C$ appearing in (11) are as given in eq.(7).
Also here $H=1+\omega^{7-p}/r^{7-p}$ and $\tilde{H}=1-\omega^{7-p}/r^{7-p}$. There are
$(4+p)$ parameters in the solution and they are $\a$, $\b$, $\omega$,
$\theta$, $\d_1$, \ldots, $\d_p$, however, there are two relations among them, one
is given in (12) and the other is $\a-\b=\d_1$. So eliminating $\a$, $\b$ we have
a $(2+p)$ parameter solution in (11). It is clear from (11) that the solution can be
made localized if the coefficients of all the terms $dx_2^2, \ldots, dx_p^2$ match
with that of the term $(-dt^2 + dx_1^2)$. This is indeed possible if $\theta=0$,
so that $F = \left(\frac{H}{\tilde H}\right)^\a$ and we set $2\d_i = - \a$, for,
$i=2,\ldots,p$. If we now define new parameters as,
\be
\tilde \a = \frac{6}{7-p} \a, \qquad \tilde{\d}_1 = \d_1 + \frac{2(p-4)}{7-p}\a
\ee
Then the solution (11) reduces to,
\bea
ds^2 &=&
(H{\tilde {H}})^{\frac{2}{7-p}}
\left(\frac{H}{\tilde H}\right)^{\frac{p+1}{8} \tilde \a}\left(dr^2 + r^2
d\Omega_{8-p}^2\right)
+ \left(\frac{H}{\tilde H}\right)^{-\frac{7-p}{8}\tilde \a} \left(-dt^2 + \sum_{i=1}^p
dx_i^2\right)\nn
e^{2\phi} &=&
\left(\frac{H}{\tilde {H}}\right)^{-a\tilde\a + 2\tilde{\delta}_1},\quad
B^{(2)} = 0, \quad a = \frac{p-3}{2}
\eea
and the parameter relation (12) reduces to,
\be
\frac{1}{2} \tilde{\d}_1^2 + \frac{1}{2} \tilde\a(\tilde\a - a \tilde{\d}_1) =
\frac{8-p}{7-p}
\ee
We recognize the solutions (14) and (15) as representing either the chargeless
D$p$-$\overline{\rm D}p$ solution or the non-BPS D$p$ solution \cite{Lu:2004ms}
depending on whether $p$ is even
or odd and whether we are considering type IIA or type IIB theory. We would like to
point out that the interpretations for the chargeless solutions (as in eq.(14)
or even the cases discussed below in eqs.(18) and (21)) are not unique. This indicates
that there are more decay processes than those considered in this paper and will be
discussed elsewhere.

Now in order to see how the solution (11) reduces to F-string solution we take the
following scaling limit
\be
\omega^{7-p} = \epsilon {\bar \omega}^{7-p}, \qquad (\a+\b) \sinh2\theta =
\epsilon^{-1}
\ee
where the parameter $\epsilon \to 0$ and ${\bar \omega}$ = fixed. Using (16) we find
from (7) $F \to {\bar H} = 1 + \frac{\bar {\omega}^{7-p}}{r^{7-p}}$ and $H$,
$\tilde H$ $\to$ 1. The solution (11) then reduces to
\bea
ds^2 &=& {\bar {H}}^{\frac{1}{4}}
\left(dr^2 + r^2
d\Omega_{8-p}^2 + \sum_{i=2}^p dx_i^2\right)
+ {\bar H}^{-\frac{3}{4}} \left(-dt^2 +
dx_1^2\right)\nn
e^{2\phi} &=&
\bar{H}^{-1},\qquad
B^{(2)}\,\,\, =\,\,\, \left(1-\bar{H}^{-1}\right) dt \wedge dx_1
\eea
We recognize (17) to be the fundamental string solution in Einstein frame delocalized
in $(p-1)$ directions. However, since this is a BPS solution, we can always make the
string solution localized by replacing $(p-1)$-dimensional source by a point source.
The harmonic function $\bar {H} = 1 + \frac{\bar{\omega}^{7-p}}{r^{7-p}}$ will then
be replaced by $\bar {H} = 1 + \frac{\bar{\omega}^{6}}{r^{6}}$. The solution (17)
will reduce to a localized F-string solution. (Note that this localization is not
a smooth operation in the parameter space. However, for the supergravity BPS solution
this is a familiar procedure when we take T-duality \cite{Breckenridge:1996tt} on it.
A similar situation arises also in the open string picture \cite{Bergman:2000xf} and
to get fully localized F-string from the decay of D2-$\overline{\rm D}$2 a 
non-pertubative technique was used there.)
This shows that the solution (11)
interpolates between D$p$-$\overline{\rm D}p$-brane (or a non-BPS D$p$-brane) solution
and the fundamental string solution. We remark that the fundamental string solutions in
both type IIA and type IIB string theory have the same forms. So, if we are in type
IIA (IIB) theory then the solution (11) interpolates between D$p$-$\overline{\rm D}p$
(non-BPS D$p$) and F-string for $p$ = even; and it interpolates between non-BPS
D$p$ (D$p$-$\overline{\rm D}p$) and F-string for $p$ = odd. We thus see that as we
move in the parameter space the interpolation has the similar effect as the open string
tachyon condensation.

Now we will see how NS5-branes can be obtained from D$p$-$\overline{\rm D}p$ solution
or non-BPS D$p$ solution. We restrict our discussion for $p=5$ and 6 only and
comment on $p<5$ later. In order to show this we start with the non-susy NS5-brane
solution delocalized in $p'$ directions. The solution in this case can be written
from (5) by putting $d=10$, $p=5$, $q=p'$, $a=-1$ (which implies $\chi=16$) as,
\bea
ds^2 &=&
F^{\frac{3}{4}} (H{\tilde {H}})^{\frac{2}{2-p'}}
\left(\frac{H}{\tilde H}\right)^{-\frac{(2\sum_{i=2}^{p'+1} \delta_i)}{(2-p')}}
\left(dr^2 + r^2
d\Omega_{3-p'}^2\right)\nn
& & + F^{-\frac{1}{4}}\left(-dt^2 + \sum_{j=1}^5 dx_j^2
\right) + F^{\frac{3}{4}}\sum_{i=2}^{p'+1}
\left(\frac{H}{\tilde H}\right)^{2\d_i} dx_{4+i}^2\nn
e^{2\phi} &=& F
\left(\frac{H}{\tilde {H}}\right)^{2\delta_1}, \quad
F_{[3]} = b {\rm Vol}(\Omega_{3-p'})\wedge dx_6 \wedge \ldots \wedge
dx_{5+p'}
\eea
Where the function $F$ is as given in (7) with $H=1+\frac{\omega^{2-p'}}{r^{2-p'}}$ and
$\tilde H=1-\frac{\omega^{2-p'}}{r^{2-p'}}$. We note that $p'$ can be 0 or 1. When it
is zero the solution (18) represents localized non-susy NS5-brane and when it is 1
it represents non-susy NS5-brane delocalized in one direction. The parameters in the
above solution are related as, $\a-\b=-\d_1$, $b=(2-p')(\a+\b)\omega^{2-p'}
\sinh2\theta$ and
\be
\frac{1}{2}\d_1^2 + \frac{1}{2} \a(\a+\d_1) + \frac{2\sum_{i<j=2}^{p'+1} \d_i \d_j}{2-p'}
= \left(1-\sum_{i=2}^{p'+1} \d_i^2\right) \frac{3-p'}{2-p'}
\ee
Note that the last term in the lhs of (19) is redundant since it is zero for both
$p'=0,1$.

Let us discuss $p'=0$ first. We note from (18) that if we put $\theta=b=0$
and define new parameters in terms of the old parameters
as $\tilde\a=\a$, $\tilde{\d}_1 = \a +\d_1$, then the solution reduces to that of
a D5-$\overline{\rm D}$5-solution or non-BPS D5 solution. Even the parameter relation
(19) reduces to $(1/2) \tilde{\d}_1^2 + (1/2) \tilde\a(\tilde\a - \tilde{\d}_1)=3/2$
i.e. that of a D5-$\overline{\rm D}$5 or non-BPS D5 solution. In order to get BPS NS5
brane solution from (18) we simply scale $\omega^2=\epsilon\,\bar{\omega}^2$ and
$(\a+\b)\sinh2\theta=\epsilon^{-1}$, where $\epsilon \to 0$ and $\bar{\omega}$ is
fixed. Then $F$ reduces to $\bar{H}=1+\frac{\bar{\omega}^2}{r^2}$ and $H$, $\tilde H$
goes to 1 and we recover BPS NS5-brane solution from (18). The form of the NSNS field
in this case is $F_{[3]} = b{\rm Vol}(\Omega_3)$. We have thus seen that
the solution (18) for $p'=0$ can be regarded as interpolating solution between
the chargeless D5-$\overline{\rm D}$5 (or non-BPS D5) solution and BPS NS5-brane
solution.

On the other hand for $p'=1$, the solution (18) can be made localized if $\theta=b=0$
and $\d_2=-\a/2$. Now defining new parameters in terms of old parameters as
$\tilde\a=2\a$ and $\tilde{\d}_1=2\a+\d_1$ we can rewrite the solution (18)
as
\bea
ds^2 &=& (H{\tilde {H}})^2
\left(\frac{H}{\tilde H}\right)^{\frac{7}{8} \tilde \a}\left(dr^2 + r^2
d\Omega_2^2\right)
+\left(\frac{H}{\tilde H}\right)^{-\frac{1}{8}\tilde\a} \left(-dt^2 + \sum_{i=1}^6
dx_i^2\right)\nn
e^{2\phi} &=&
\left(\frac{H}{\tilde {H}}\right)^{-\frac{3}{2}\tilde\a + 2\tilde{\delta}_1},
\qquad F_{[3]} = 0
\eea
The parameter relation (19) can also be rewritten as $(1/2) \tilde{\d}_1^2
+ (1/2) \tilde\a(\tilde\a-(3/2)\tilde{\d}_1) = 2$. The above solution represents
precisely the D6-$\overline{\rm D}$6 (or non-BPS D6) brane solution. Again in order
to get BPS NS5-brane solution from (18), we do the same scaling of $\omega$
and $\theta$ as before. But, since in this case the harmonic functions are that
of a 6-brane involving $(1/r)$ and not $(1/r^2)$, we get a delocalized NS5-brane
where the form-field is given as $F_{[3]} = b {\rm Vol}(\Omega_2)\wedge dx_6$.
But since this solution is BPS, one can make it localized
as mentioned before. This shows that the solution (18) for $p'=1$ can
indeed be regarded as interpolating solution between D6-$\overline{\rm D}$6 (or
non-BPS D6) brane solution and BPS NS5-brane solution.

For other values of $p<5$, it is not clear how the above procedure will work.
But one can start with the non-isotropic, non-susy NS5-brane solution and try to
make it D$p$-$\overline{\rm D}p$ (or non BPS D$p$) solution as we have done above.
But it can be easily checked that the resulting solution can not be made localized
D$p$-$\overline{\rm D}p$ (or non-BPS D$p$) by only adjusting the parameters. If we
look at only the linear part of the solution then the solution will look like
localized solution. The meaning of this is not entirely clear to us.

Next we will show how D$p$-branes (for $p\leq 5$) can be obtained as the
decay product of
NS5-$\overline{\rm NS}$5. In supergravity we describe the process
with a solution which interpolates between NS5-$\overline{\rm NS}$5 and D$p$
brane solutions. For this purpose we write non-susy D$p$-brane solution
delocalized in $(5-p)$-directions from (5) and (6) by putting $d=10$, $p=p$,
$q=5-p$, $a=(p-3)/2$ (which implies $\chi = 32/(7-p)$) as,
\bea
ds^2 &=&
F^{\frac{p+1}{8}} (H{\tilde {H}})
\left(\frac{H}{\tilde H}\right)^{-\sum_{i=2}^{6-p} \delta_{i}}\left(dr^2 + r^2
d\Omega_3^2\right)\nn
&& + F^{-\frac{7-p}{8}}\left(-dt^2 + \sum_{i=1}^p dx_i^2
\right)
+ F^{\frac{p+1}{8}}\sum_{i=2}^{6-p}
\left(\frac{H}{\tilde H}\right)^{2\d_i} dx_{i+p-1}^2\nn
e^{2\phi} &=& F^{-\frac{p-3}{2}}
\left(\frac{H}{\tilde {H}}\right)^{2\delta_1}\nn
A^{(p+1)} &=&  \frac{\sinh2\theta}{2} \left(\frac{C}{F}\right) dt \wedge dx_1 \wedge
\ldots \wedge dx_p
\eea
The parameters satisfy,
\be
\frac{1}{2}\d_1^2 + \frac{1}{2} \a\left(\a-a\d_1\right) +
\sum_{i<j=2}^{6-p} \d_i \d_j
= \left(1-\sum_{i=2}^{6-p} \d_i^2\right) \frac{3}{2}
\ee
Where in the above the functions $F$, $C$ are as defined before,
but $H,\,\tilde H$ have
the forms $H=1+\frac{\omega^2}{r^2}$ and $\tilde H = 1 - \frac{\omega^2}{r^2}$. The
above solution can be made localized if we put $\theta=0$ and $-2\delta_i = \alpha$,
for $i=2,\ldots,(6-p)$. Defining new parameters as $\tilde\a=(7-p)\a/2$ and
$\tilde\d_1 = \d_1 - \a$, the solution (21) reduces to,
\bea
ds^2 &=& (H{\tilde {H}})
\left(\frac{H}{\tilde H}\right)^{\frac{3}{4}\tilde\a}\left(dr^2 + r^2
d\Omega_{3}^2\right)
+ \left(\frac{H}{\tilde H}\right)^{-\frac{1}{4}\tilde\a} \left(-dt^2 + \sum_{i=1}^5
dx_i^2\right)\nn
e^{2\phi} &=&
\left(\frac{H}{\tilde {H}}\right)^{\tilde\a + 2\tilde{\delta}_1},\qquad
A^{(p+1)} = 0
\eea
and the parameter relation (22) takes the form $(1/2)\tilde{\d}_1^2 + (1/2) \tilde\a
(\tilde\a+\tilde{\d}_1) = 3/2$. This is precisely the chargeless NS5-$\overline{\rm NS}$5
brane solution \cite{Lu:2004ms}. Now in order to see how the solution (21) reduces
to the BPS D$p$-brane
solution, we scale the parameters $\omega$ and $\theta$ exactly as before and send
$H,\, \tilde H \to 1$ and $F \to {\bar H} = 1+ \frac{\bar{\omega}^2}{r^2}$. $A^{(p+1)}$
in (21) reduces to $(1-{\bar H}^{-1})dt\wedge dx_1\wedge \ldots \wedge dx_p$.
The solution then
reduces precisely to the D$p$-brane solution delocalized in $(5-p)$ directions. However,
it can be easily localized by replacing the $(5-p)$-dimensional source by a point
source as we mentioned earlier. This therefore shows that the delocalized solution
(21) can indeed be regarded as the interpolating solution between
the chargeless NS5-$\overline{\rm NS}$5-brane and the D$p$-brane. From the world-volume
point of view this interpolation means D$p$-branes appear as a decay product of
NS5-$\overline{\rm NS}$5-brane.

We have thus seen how the various delocalized, non-susy $p$-brane solutions of
type II supergravities can be regarded as interpolating solutions between
D$p$-$\overline{\rm D}p$ (or non-BPS D$p$)-brane and F-string, between
D$p$-$\overline{\rm D}p$ (or non-BPS D$p$)-brane and NS5-brane and
NS5-$\overline{\rm NS}$5-brane
and D$p$-brane solutions, by adjusting and scaling the parameters of the solutions.
The open string description of some of these processes were well-understood in
terms of tachyon condensation. We here obtain a closed string or supergravity picture
of these processes.

\vspace{.5cm}

\noindent {\bf Acknowledgements}

\vspace{2pt}

One of us (JXL) acknowledges the support by grants from the
Chinese Academy of Sciences and a grant from the NSF of China with
grant no. 90303002.

\medskip


\end{document}